\documentclass[a4paper, 11pt]{article}

\usepackage{jheppub}
\usepackage{graphicx}
\usepackage{here}

\title{
Constraints on the non-standard
interaction in propagation
from atmospheric neutrinos
}

\author{Shinya Fukasawa,}

\author{Osamu Yasuda}
\affiliation{Department of Physics, Tokyo Metropolitan University,
Minami-Osawa, Hachioji, Tokyo 192-0397, Japan}

\abstract{
The sensitivity of the atmospheric neutrino
experiments to the non-standard flavor-dependent
interaction
in neutrino propagation is studied under the
assumption that the only nonvanishing components
of the non-standard matter effect are
the electron
and tau neutrino components $\epsilon_{ee}$,
$\epsilon_{e\tau}$, $\epsilon_{\tau\tau}$
and that the tau-tau component satisfies
the constraint
$\epsilon_{\tau\tau}=|\epsilon_{e\tau}|^2/(1+\epsilon_{ee})$
which is suggested from the high energy
behavior for atmospheric neutrino data.
It is shown that the Superkamiokande (SK)
data for 4438 days constrains
$|\tan\beta|\equiv|\epsilon_{e\tau}/(1+\epsilon_{ee})|\lesssim 0.8$
at 2.5$\sigma$ (98.8\%) CL
whereas the future Hyperkamiokande experiment
for the same period of time as SK will constrain as
$|\tan\beta|\lesssim 0.3$
at 2.5$\sigma$CL
from the energy rate analysis and
the energy spectrum analysis will give even tighter bounds
on $\epsilon_{ee}$ and $|\epsilon_{e\tau}|$.}

\begin{document}

\maketitle

\section{introduction}

From the experiments with solar, atmospheric,
reactor and accelerator neutrinos
it is now established that neutrinos have masses
and mixings\,\cite{Agashe:2014kda}.
Neutrino oscillations in the standard three-flavor scheme
are described by three mixing angles,
$\theta_{12}$, $\theta_{13}$, $\theta_{23}$, one CP phase $\delta$,
and two independent mass-squared differences, $\Delta m^2_{21}$ and
$\Delta m^2_{31}$.  The sets of the parameters
($\Delta m^2_{21}$, $\theta_{12}$) and
($|\Delta m^2_{31}|$, $\theta_{23}$)
were determined by
the solar neutrino experiments and the KamLAND experiment,
and by atmospheric and long baseline
neutrino experiments, respectively.
$\theta_{13}$ was determined by
the reactor experiments
and the long baseline
experiments\,\cite{Agashe:2014kda}.
The only oscillation parameters which are
still undetermined are
the value of the CP phase $\delta$ and the sign of
$\Delta m^2_{31}$ (the mass hierarchy).
In the future neutrino long-baseline experiments
with intense neutrino beams 
the sign of $\Delta m^2_{31}$ and
$\delta$ are expected to be determined\,\cite{Abe:2014oxa,Adams:2013qkq}.
As in the case of B factories\,\cite{belle:0000,babar:0000},
such high precision measurements will enable us to
search for deviation from the standard three-flavor oscillations
(see, e.g., Ref.\,\cite{Bandyopadhyay:2007kx}).
Among such possibilities,
in this paper, we will discuss
the effective non-standard neutral current
flavor-dependent neutrino interaction with
matter\,\cite{Wolfenstein:1977ue,Guzzo:1991hi,Roulet:1991sm} given by
\begin{eqnarray}
{\cal L}_{\mbox{\rm\scriptsize eff}}^{\mbox{\tiny{\rm NSI}}} 
=-2\sqrt{2}\, \epsilon_{\alpha\beta}^{fP} G_F
(\overline{\nu}_\alpha \gamma_\mu P_L \nu_\beta)\,
(\overline{f} \gamma^\mu P f'),
\label{NSIop}
\end{eqnarray}
where $f$ and $f'$ stand for fermions (the only relevant
ones are electrons, u and d quarks),
$G_F$ is the Fermi coupling constant, and $P$ stands for
a projection operator that is either
$P_L\equiv (1-\gamma_5)/2$ or $P_R\equiv (1+\gamma_5)/2$.
If the interaction (\ref{NSIop}) exists, then
the standard matter effect\,\cite{Wolfenstein:1977ue,Mikheev:1986gs}
is modified.
We will discuss atmospheric neutrinos
which go through the Earth, so we make an approximation
that the number densities of electrons ($N_e$),
protons, and neutrons are equal.\footnote{
This assumption is not valid in other environments, e.g., in the Sun.}
Defining
$\epsilon_{\alpha\beta}
\equiv \sum_{P}
\left(
\epsilon_{\alpha\beta}^{eP}
+ 3 \epsilon_{\alpha\beta}^{uP}
+ 3 \epsilon_{\alpha\beta}^{dP}
\right)$,
the hermitian $3 \times3 $ matrix of the matter potential becomes
\begin{eqnarray}
{\cal A}\equiv A\left(
\begin{array}{ccc}
1+ \epsilon_{ee} & \epsilon_{e\mu} & \epsilon_{e\tau}\\
\epsilon_{\mu e} & \epsilon_{\mu\mu} & \epsilon_{\mu\tau}\\
\epsilon_{\tau e} & \epsilon_{\tau\mu} & \epsilon_{\tau\tau}
\end{array}
\right),
\label{matter-np}
\end{eqnarray}
where $A\equiv\sqrt{2}G_FN_e$ stands for the matter effect
due to the charged current interaction in the standard case.
With this matter potential, the Dirac equation for
neutrinos in matter becomes
\begin{eqnarray}
i {d \over dx} \left( \begin{array}{c} \nu_e(x) \\ \nu_{\mu}(x) \\ 
\nu_{\tau}(x)
\end{array} \right) = 
\left[  U {\rm diag} \left(0, \Delta E_{21}, \Delta E_{31}
\right)  U^{-1}
+{\cal A}\right]
\left( \begin{array}{c}
\nu_e(x) \\ \nu_{\mu}(x) \\ \nu_{\tau}(x)
\end{array} \right),
\label{eqn:sch}
\end{eqnarray}
where $U$ is the leptonic mixing matrix defined by
\begin{eqnarray}
U&\equiv&\left(
\begin{array}{ccc}
c_{12}c_{13} & s_{12}c_{13} &  s_{13}e^{-i\delta}\cr
-s_{12}c_{23}-c_{12}s_{23}s_{13}e^{i\delta} & 
c_{12}c_{23}-s_{12}s_{23}s_{13}e^{i\delta} & s_{23}c_{13}\cr
s_{12}s_{23}-c_{12}c_{23}s_{13}e^{i\delta} & 
-c_{12}s_{23}-s_{12}c_{23}s_{13}e^{i\delta} & c_{23}c_{13}
\end{array}
\right),
\label{eqn:mns3}
\end{eqnarray}
and $\Delta E_{jk}\equiv\Delta m_{jk}^2/2E\equiv (m_j^2-m_k^2)/2E$,
$c_{jk}\equiv\cos\theta_{jk}$, $s_{jk}\equiv\sin\theta_{jk}$.

Constraints on $\epsilon_{\alpha\beta}$
from various neutrino experiments have been discussed
in Refs.\,\cite{Fornengo:2001pm,Berezhiani:2001rs,Davidson:2003ha,GonzalezGarcia:2004wg,Miranda:2004nb,Barranco:2005ps,Barranco:2007ej,Bolanos:2008km,Escrihuela:2009up}.
Since the coefficients $\epsilon_{\alpha\beta}$ in Eq.\,(\ref{matter-np})
are given by
$\epsilon_{\alpha\beta}\sim
\epsilon^e_{\alpha\beta}
+3\epsilon^u_{\alpha\beta}
+3\epsilon^d_{\alpha\beta}$,
considering the constraints by 
Refs.\,\cite{Fornengo:2001pm,Berezhiani:2001rs,Davidson:2003ha,GonzalezGarcia:2004wg,Miranda:2004nb,Barranco:2005ps,Barranco:2007ej,Bolanos:2008km,Escrihuela:2009up},
we have the following limits\,\cite{Biggio:2009nt} at 90\%CL:
\begin{eqnarray}
\left(
\begin{array}{ccc}
|\epsilon_{ee}| < 4\times 10^0 & \quad|\epsilon_{e\mu}| < 3\times 10^{-1}
& \quad|\epsilon_{e\tau}| < 3\times 10^{0\ }\\
& \quad |\epsilon_{\mu\mu}| < 7\times 10^{-2}
& \quad|\epsilon_{\mu\tau}| < 3\times 10^{-1}\\
& & \quad|\epsilon_{\tau\tau}| < 2\times 10^{1\ }
\end{array}
\right).
\label{epsilon-m}
\end{eqnarray}
From Eq.\,(\ref{epsilon-m}) we observe that the bounds on 
$\epsilon_{ee}$, $\epsilon_{e\tau}$ and $\epsilon_{\tau\tau}$ are
much weaker than those on $\epsilon_{\alpha\mu}~(\alpha=e,\mu,\tau)$.

On the other hand, the non-standard interaction (NSI)
with components $\epsilon_{\alpha\beta}~(\alpha,\beta=e,\tau)$
must be consistent with
the high-energy atmospheric neutrino data.
It was pointed out in Refs.\,\cite{Friedland:2004ah,Friedland:2005vy}
that the relation
\begin{eqnarray}
|\epsilon_{e\tau}|^2
\simeq \epsilon_{\tau\tau} \left( 1 + \epsilon_{ee} \right),
\label{atm}
\end{eqnarray}
should hold for the matter potential (\ref{matter-np}) to be consistent with
the high-energy atmospheric neutrino data, which suggest the
behavior of the disappearance oscillation probability
\begin{eqnarray}
1-P(\nu_\mu\rightarrow\nu_\mu)\sim
\sin^22\theta_{\text{atm}}
\sin^2\left(\frac{\Delta m^2_{\text{atm}}L}{4E}\right)\,
\propto\, \frac{1}{E^2}\,,
\label{atm-he}
\end{eqnarray}
where $\sin^22\theta_{\text{atm}}$ and
$\Delta m^2_{\text{atm}}$ are the oscillation
parameters in the two-flavor formalism.
In Ref.\,\cite{Oki:2010uc} it was shown that,
in the high-energy behavior of the
disappearance oscillation probability
\begin{eqnarray}
1-P(\nu_\mu\rightarrow\nu_\mu)\simeq
c_0 + \frac{c_1}{E} + {\cal O}\left(\frac{1}{E^2}\right),
\label{expansion}
\end{eqnarray}
in the presence of the matter potential (\ref{matter-np}),
$|c_0| \ll 1$ and $|c_1| \ll 1$ imply
$\epsilon_{e\mu}\simeq\epsilon_{\mu\mu}\simeq\epsilon_{\mu\tau}\simeq0$
and
$\epsilon_{\tau\tau}\simeq 
|\epsilon_{e\tau}|^2/ \left( 1 + \epsilon_{ee} \right)$.

Taking into account the various constraints described above,
in the present paper we take the ansatz
\begin{eqnarray}
{\cal A}= A\left(
\begin{array}{ccc}
1+ \epsilon_{ee}~~ & 0 & \epsilon_{e\tau}\\
0 & 0 & 0\\
\epsilon_{e\tau}^\ast & 0 & ~~|\epsilon_{e\tau}|^2/(1 + \epsilon_{ee})
\end{array}
\right),
\label{ansatz}
\end{eqnarray}
and analyze the sensitivity to
the parameters $\epsilon_{\alpha\beta}~(\alpha,\beta=e,\tau)$
of the atmospheric neutrino experiment at
Superkamiokande and the future Hyperkamiokande (HK)
facility\,\cite{Abe:2011ts}.

The constraints on $\epsilon_{ee}$ and $\epsilon_{e\tau}$
from the atmospheric neutrino has been discussed in
Refs~\cite{GonzalezGarcia:2011my, Mitsuka:2011ty, Gonzalez-Garcia:2013usa}
with the ansatz different from ours.

The effect of the non-standard interaction in propagation for solar neutrinos
has also been discussed in Refs.\,\cite{Friedland:2004pp,Miranda:2004nb,Bolanos:2008km,Escrihuela:2009up,Palazzo:2009rb}, and
Refs.\,\cite{Escrihuela:2009up} and\,\cite{Palazzo:2009rb}
give a constraint $-0.06<\epsilon_{e\tau}^{dV}\sin\theta_{23}< 0.41$
(at 90\%CL) and 
$|\epsilon_{e\tau}^{dV}|\lesssim 0.4$ (at $\Delta\chi^2=4$ for 2 d.o.f.),
respectively.

The sensitivity of the ongoing long-baseline experiments to the non-standard interaction in
propagation was studied for MINOS in
Refs.\,\cite{Friedland:2006pi,Yasuda:2007tx,Sugiyama:2007ub,Blennow:2007pu,Coelho:2012bp}, and for
OPERA in
Refs.\,\cite{EstebanPretel:2008qi,Blennow:2008ym}.
As for the future long-baseline experiments,
the sensitivity of the reactor and super-beam experiments was
discussed in Ref.\,\cite{Kopp:2007ne},
that of the T2KK experiment
was studied in Refs.\,\cite{Ribeiro:2007jq,Oki:2010uc},
and that of neutrino
factories\,\cite{Bandyopadhyay:2007kx} was discussed
by many
authors\,\cite{Gago:2001xg,Ota:2001pw,Huber:2001de,Campanelli:2002cc,Ribeiro:2007ud,Kopp:2008ds,Gago:2009ij,Meloni:2009cg}.

The paper is organized as follows.
In sect.\,\ref{atmospheric-sk}, we analyze
the SK atmospheric neutrino data and give
the constraints
on the parameters $\epsilon_{\alpha\beta}~(\alpha,\beta=e,\tau)$
from the SK atmospheric neutrino data.
In sect.\,\ref{atmospheric-hk}, we discuss
the sensitivity to $\epsilon_{\alpha\beta}~(\alpha,\beta=e,\tau)$.
of the future Hyperkamiokande
atmospheric neutrino experiment
In sect.\,\ref{conclusions}, we draw our conclusions.

\section{The constraint of the Superkamiokande atmospheric
neutrino experiment on $\epsilon_{ee}$ and $|\epsilon_{e\tau}|$
\label{atmospheric-sk}}

In this section we discuss the constraint of the
SK atmospheric neutrino experiment on
the non-standard interaction in propagation with the ansatz
(\ref{ansatz}).
The independent degrees of freedom
in addition to those in the standard
oscillation scenario are
$\epsilon_{ee}$, $|\epsilon_{e\tau}|$
and $\mbox{\rm arg}(\epsilon_{e\tau})$.

The SK atmospheric neutrino data we analyze here is
those in Ref.\,\cite{Abe:2014gda} for 4438
days.  In Ref.\,\cite{Abe:2014gda}, the contained events,
the partially contained events and the upward going $\mu$
events are divided into a few categories.
Since we have been unable to reproduce all their results of the Monte Carlo
simulation, we have combined the two sub-GeV $\mu$-like
data set in one, the two multi-GeV e-like in one,
the two partially contained event data set and
the multi-GeV $\mu$-like in one, and
the three upward going $\mu$ in one.
Ref.\,\cite{Abe:2014gda} gives information on
the ten zenith angle bins, while
that on the energy bins is not given,
so we perform analysis with 
the ten zenith angle bins
and one energy bin, i.e., we perform
the rate analysis as far as the energy is concerned.

The analysis was performed with the codes which were
used in Refs.\,\cite{Foot:1998iw,Yasuda:1998mh,Yasuda:2000de}.
$\chi^2$ is defined as
\begin{eqnarray}
\chi^2=
\min_{\theta_{23},|\Delta m^2_{32}|,\delta, \mbox{\small\rm arg}(\epsilon_{e\tau})}
\left(
\chi_{\rm sub-GeV}^2+\chi_{\rm multi-GeV}^2
+\chi_{\rm upward}^2\right).
\label{eqn:chi}
\end{eqnarray}
In eq.\,(\ref{eqn:chi}) $\chi^2$ for the sub-GeV, multi-GeV, and upward going
$\mu$ events are defined by
\hglue -1cm
\begin{eqnarray}
\hspace*{-40mm}
\displaystyle\chi_{\rm sub-GeV}^2
&=&
\min_{\alpha_s,\beta's}\left[
\frac{\beta_{s1}^2}{\sigma_{\beta s1}^2}
+\frac{\beta_{s2}^2}{\sigma_{\beta s2}^2}\right.
+\sum_{j=1}^{10}\left\{
\frac{1}{n_j^s(e)}
\left[ \alpha_s \left(1-{\beta_{s1} \over 2}+{\beta_{s2}
\over 2}\right)N_j^s(\nu_e\to\nu_e)
\right.\right.
\nonumber\\
&{\ }&\quad\quad+ \alpha_s \left(1+{\beta_{s1} \over 2}+{\beta_{s2}
\over 2}\right)N_j^s(\nu_\mu\to\nu_e)
\nonumber\\
&{\ }&\quad\quad+ \alpha_s \left(1-{\beta_{s1} \over 2}-{\beta_{s2}
\over 2}\right)N_j^s(\bar{\nu}_e\to\bar{\nu}_e)
\nonumber\\
&{\ }&\quad\quad\left.
+ \alpha_s \left(1+{\beta_{s1} \over 2}-{\beta_{s2}
\over 2}\right)N_j^s(\bar{\nu}_\mu\to\bar{\nu}_e)
-n_j^s(e)\right]^2
\nonumber\\
&{\ }&\quad+\frac{1}{n_j^s(\mu)}
\left[ 
\alpha_s \left(1-{\beta_{s1} \over 2}+{\beta_{s2} 
\over 2}\right)N_j^s(\nu_e\to\nu_\mu)\right.
\nonumber\\
&{\ }&\quad\quad+\alpha_s \left(1+{\beta_{s1} \over 2}+{\beta_{s2} 
\over 2}\right)N_j^s(\nu_\mu\to\nu_\mu)
\nonumber\\
&{\ }&\quad\quad+\alpha_s \left(1-{\beta_{s1} \over 2}-{\beta_{s2} 
\over 2}\right)N_j^s(\bar{\nu}_e\to\bar{\nu}_\mu)
\nonumber\\
&{\ }&\quad\quad\left.\left.\left.+\alpha_s \left(1+{\beta_{s1} \over 2}-{\beta_{s2} 
\over 2}\right)N_j^s(\bar{\nu}_\mu\to\bar{\nu}_\mu)
-n_j^s(\mu)\right]^2
\right\}\right],
\label{eqn:chi-sub1}
\\
\nonumber\\
\displaystyle\chi_{\rm multi-GeV}^2
&=&
\min_{\alpha_m,\beta's}\left[
\frac{\beta_{m1}^2}{\sigma_{\beta m1}^2}
+\frac{\beta_{m2}^2}{\sigma_{\beta m2}^2}
+\sum_{j=1}^{10}\left\{
\frac{1}{n_j^m(e)}
\left[ \alpha_s \left(1-{\beta_{m1} \over 2}+{\beta_{m2}
\over 2}\right)N_j^m(\nu_e\to\nu_e)\right.
\right.\right.
\nonumber\\
&{\ }&\quad\quad
+ \alpha_s \left(1+{\beta_{m1} \over 2}+{\beta_{m2}
\over 2}\right)N_j^m(\nu_\mu\to\nu_e)
\nonumber\\
&{\ }&\quad\quad+ \alpha_s \left(1-{\beta_{m1} \over 2}-{\beta_{m2}
\over 2}\right)N_j^m(\bar{\nu}_e\to\bar{\nu}_e)
\nonumber\\
&{\ }&\quad\quad\left.
+ \alpha_s \left(1+{\beta_{m1} \over 2}-{\beta_{m2}
\over 2}\right)N_j^m(\bar{\nu}_\mu\to\bar{\nu}_e)
-n_j^m(e)\right]^2
\nonumber\\
&{\ }&\quad+\frac{1}{n_j^m(\mu)}
\left[ 
\alpha_s \left(1-{\beta_{m1} \over 2}+{\beta_{m2} 
\over 2}\right)N_j^m(\nu_e\to\nu_\mu)\right.
\nonumber\\
&{\ }&\quad\quad+\alpha_s \left(1+{\beta_{m1} \over 2}+{\beta_{m2} 
\over 2}\right)N_j^m(\nu_\mu\to\nu_\mu)
\nonumber\\
&{\ }&\quad\quad+\alpha_s \left(1-{\beta_{m1} \over 2}-{\beta_{m2} 
\over 2}\right)N_j^m(\bar{\nu}_e\to\bar{\nu}_\mu)
\nonumber\\
&{\ }&\quad\quad\left.\left.\left.+\alpha_s \left(1+{\beta_{m1} \over 2}-{\beta_{m2} 
\over 2}\right)N_j^m(\bar{\nu}_\mu\to\bar{\nu}_\mu)
-n_j^m(\mu)\right]^2
\right\}\right],
\label{eqn:chi-multi1}
\\
\nonumber\\
\displaystyle\chi_{\rm upward}^2
&=&
\min_{\alpha_u}\left\{
\frac{\alpha_u^2}{\sigma_{\alpha}^2}
+\sum_{j=1}^{10}
\frac{1}{n_j^u(\mu)}
\left[ 
\alpha_u N_j^u(\nu_e\to\nu_\mu)
+\alpha_u N_j^u(\nu_\mu\to\nu_\mu)\right.\right.
\nonumber\\
&{\ }&\qquad\left.\left.+\alpha_u N_j^u(\bar{\nu}_e\to\bar{\nu}_\mu)
+\alpha_u N_j^u(\bar{\nu}_\mu\to\bar{\nu}_\mu)
-n_j^u(\mu)\right]^2\right\}.
\nonumber
\end{eqnarray}
The summation on $j$ runs
over the ten zenith angle bins for each $\chi^2$,
$n_j^a(\alpha)$ ($a$=s, m, u; $\alpha=e,\mu$) stands
for the neutrino {\it and} antineutrino data
of the numbers of the sub-GeV, multi-GeV, and upward going
$\mu$ events,
$N_j^a(\nu_\alpha\to\nu_\beta)$
($N_j^a(\bar{\nu}_\alpha\to\bar{\nu}_\beta)$)
 stands for the
theoretical prediction for the number of
$\ell_\beta$-like events ($\ell_\beta=e,\mu$)
which is produced from $\nu_\beta$
($\bar{\nu}_\beta$)
that originates from $\nu_\alpha$
($\bar{\nu}_\alpha$)
through the oscillation process
$\nu_\alpha\to\nu_\beta$
($\bar{\nu}_\alpha\to\bar{\nu}_\beta$), and it is expressed as
the product of the oscillation probability
$P(\nu_\alpha\to\nu_\beta)$
($P(\bar{\nu}_\alpha\to\bar{\nu}_\beta)$), the flux
$F(\nu_\alpha)$ ($F(\bar{\nu}_\alpha)$), the cross section, the number
of the targets and the detection efficiency.
$\alpha_a~(a=s, m, u)$ stands for the uncertainty
in the overall flux normalization for the sub-GeV,
multi-GeV, and upward going $\mu$ events,
$\beta_{a1}$ ($\beta_{a2}$) stands for the uncertainty
in the relative normalization between
$\nu_e$ - $\nu_\mu$ flux ($\nu$ - $\bar{\nu}$ flux)
for the sub-GeV ($a=s$) and multi-GeV ($a=m$) events,
respectively.  It is understood that
$\chi^2$ is minimized with respect to
$\alpha_s$, $\beta_{sk}$ ($k=1,2$), $\alpha_m$,
$\beta_{mk}$ ($k=1,2$), $\alpha_u$.
We have put the systematic errors
\begin{eqnarray}
\sigma_{\beta s1}=\sigma_{\beta m1} = 0.03,
\sigma_{\beta s2}=\sigma_{\beta m2} = 0.05,
\sigma_{\alpha} = 0.2
\label{sys1}
\end{eqnarray}
and we have assumed that
$\alpha_s$ and $\alpha_m$ for
the contained events are free parameters as in Ref.\,\cite{Ashie:2005ik}.
We have omitted the other uncertainties, such as the $E_\nu$ spectral index,
the relative normalization between PC and FC and up-down
correlation, etc., for simplicity.
In Eq. (\ref{eqn:chi}) the sum of each $\chi^2$ is optimized
with respect the mixing angle $\theta_{23}$,
the mass squared difference $|\Delta m^2_{32}|$,
the Dirac CP phase $\delta$
and the phase arg($\epsilon_{e\tau}$)
of the parameter $\epsilon_{e\tau}$.
The other oscillation parameters
give little effect on $\chi^2$, so
we have fixed them as $\sin^22\theta_{12}=0.86$,
$\sin^22\theta_{13}=0.1$ and
$\Delta m^2_{21}=7.6\times 10^{-5}$eV$^2$.

\begin{figure}[H]
\includegraphics[scale=0.4]{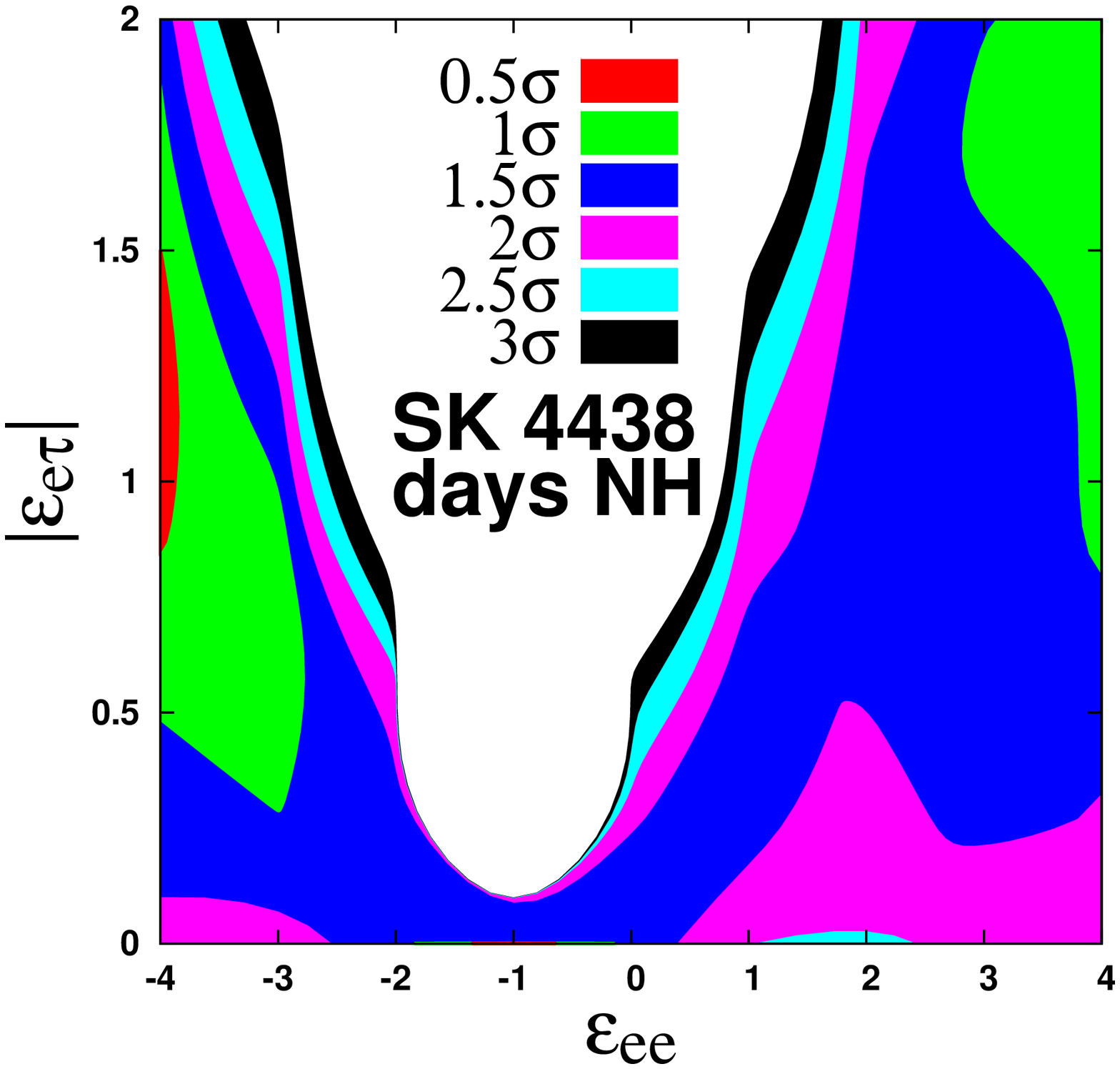}
\includegraphics[scale=0.4]{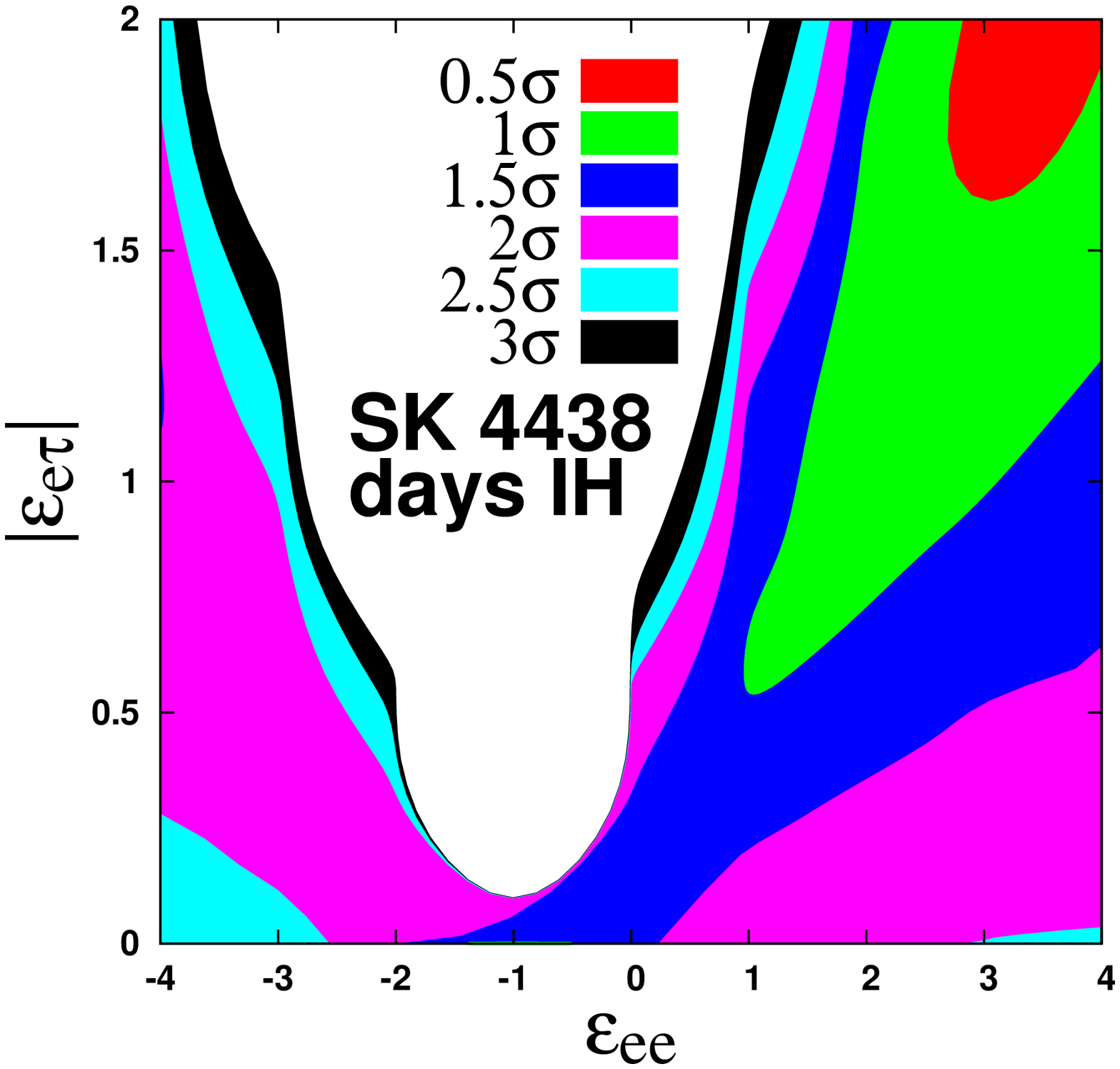}
\vspace*{-5mm}
\caption{
The allowed region
in the ($\epsilon_{ee}$, $|\epsilon_{e\tau}|$) plane
from the SK atmospheric neutrino data for a
normal mass hierarchy (left panel) and for an inverted mass
hierarchy (right panel).
}
\label{fig:fig1}
\end{figure}

The result for the Superkamiokande data
for 4438 days is given in Fig.\,\ref{fig:fig1}.
The best-fit point for the normal (inverted) hierarchy is
($\epsilon_{ee}$, $|\epsilon_{e\tau}|$) = (-1.0, 0.0)
((3.0, 1.7))
and the value of $\chi^2$ at this point is
79.0 (78.6) for 50 degrees of freedom, and
goodness of fit is 2.8 (2.7) $\sigma$CL,
respectively.
The best-fit point is different from the
standard case 
($\epsilon_{ee}$, $|\epsilon_{e\tau}|$) = (0, 0),
and this may be not only because we have been unable to reproduce
the Monte Carlo
simulation by the Superkamiokande group,
but also because we use only the information
on the energy rate and the sensitivity to NSI
is lost due to the destructive phenomena
between the lower and higher energy bins
(See the discussions in subsect.\,\ref{hk-std}).
The difference of the value of $\chi^2$ for the
standard case and that for the best-fit point
for the normal (inverted) hierarchy
is $\Delta\chi^2=2.7$ (3.4) for 2 degrees of freedom
(1.1 $\sigma$CL (1.3 $\sigma$CL)), respectively,
and the standard case is certainly acceptable
for the both mass hierarchies in our analysis.
From the Fig.\,\ref{fig:fig1} we can read off
the allowed region for
$|\tan\beta|\equiv|\epsilon_{e\tau}|/|1+\epsilon_{ee}|$,
and we conclude that
the allowed region for $|\tan\beta|$
is approximately
\begin{eqnarray}
|\tan\beta|\equiv\frac{|\epsilon_{e\tau}|}{|1+\epsilon_{ee}|}\lesssim 0.8
\qquad\mbox{\rm at 2.5$\sigma$CL}.
\nonumber
\end{eqnarray}

\section{Sensitivity of the Hyperkamiokande atmospheric
neutrino experiment to $\epsilon_{ee}$ and $|\epsilon_{e\tau}|$
\label{atmospheric-hk}}

In this section we discuss the potential sensitivity of HK to
$\epsilon_{ee}$ and $|\epsilon_{e\tau}|$.
Here we assume for simplicity that the
the Hyperkamiokande detector has
the same detection efficiencies
as those of SK, and that
the fiducial volume of HK
is twenty times as large as that of SK.
Since HK is a future experiment, the simulated
numbers of events are used as ``the experimental data'',
and we vary $\epsilon_{ee}$ and $\epsilon_{e\tau}$
as well as the standard oscillation parameters
trying to fit to ``the experimental data''.
Here we perform an analysis
on the assumption that we know the mass
hierarchy, because
some hint on the mass hierarchy is
expected to be available at some confidence level
by the time HK will
accumulate the atmospheric neutrino data for twenty years.

Since ``the experimental data''
are the simulated numbers of events,
we can perform a energy spectrum analysis,
assuming that the detection efficiency etc.
are all equal among neutrinos and antineutrinos.
Before we study the sensitivity to NSI,
as a benchmark of our analysis,
we have investigated the significance
of the wrong mass hierarchy with our code,
assuming the standard oscillation scenario
and using different numbers of the energy bins.
By comparing our result with the one in
Ref.\,\cite{Abe:2011ts},
we have found that our analysis on the
mass hierarchy gives a result similar to
that in Ref.\,\cite{Abe:2011ts},
when we work with two
energy bins in the contained events
(the sub-GeV and multi-GeV events)
and the systematic errors which are
slightly different from those in Ref.\,\cite{Abe:2011ts}.
We have checked that the sensitivity to NSI
is not affected significantly by changing the systematic errors.
As for the upward going $\mu$ events,
since our ansatz (\ref{ansatz}) is
taken in such a way that the oscillation
probability with
$\epsilon_{\alpha\beta}~(\alpha, \beta = e, \tau)$
approaches to the one with the standard 
scenario in the high energy limit,
the upward going $\mu$ events are
expected to give a small contribution
to the significance of NSI.
So in the case of the energy spectrum analysis
we will work with two energy bins
in the contained events and a single energy
bin in the upward going $\mu$ events.

\subsection{The case with the standard oscillation
scenario
\label{hk-std}}

First of all, let us discuss the case where
``the experimental data'' is the one
obtained with the standard oscillation
scenario.  The values of the
oscillation parameters which are used
to obtain ``the experimental data''
are the following\,\footnote{
To distinguish the oscillation parameters
for the ``the experimental data''
($n_{Aj}^a(\ell)~(j=1,\cdots,10; A=L,H; a=s,m; \ell=e,\mu)$,
etc.) and those for the numbers of events
($N_{Aj}^a(\nu_\alpha\to\nu_\beta)~(j=1,\cdots,10; A=L,H;
a=s,m; \alpha,\beta=e,\mu)$,
etc.) for fitting, the parameters with a bar denote
those for ``the experimental data'',
whereas those without a bar denote
the parameters for the numbers of events for fitting.}:
\begin{eqnarray}
\Delta \bar{m}^2_{31}=2.5\times 10^{-3}\mbox{\rm eV}^2,
\sin^2\bar{\theta}_{23}=0.5,
\bar{\delta}=0,
\nonumber\\
\sin^22\bar{\theta}_{12}=0.86,
\sin^22\bar{\theta}_{13}=0.1,
\Delta \bar{m}^2_{21}=7.6\times 10^{-5}\mbox{\rm eV}^2.
\label{ref-value}
\end{eqnarray}
As in the case of the analysis of the SK data,
we vary the oscillation parameters
$\theta_{23}$, $|\Delta m^2_{32}|$,
$\delta$ and arg($\epsilon_{e\tau}$)
while fixing the other oscillation parameters
$\sin^22\theta_{12}=0.86$,
$\sin^22\theta_{13}=0.1$ and
$\Delta m^2_{21}=7.6\times 10^{-5}$eV$^2$.

In the energy rate analysis,
$\chi^2$ is the same as (\ref{eqn:chi})
where the numbers of events are
calculated with the standard oscillation
scenario with $\bar{\theta}_{jk}$,
$\Delta \bar{m}^2_{jk}$ and $\bar{\delta}$
given by Eq.\,(\ref{ref-value}), and
we have assumed that all the systematic errors
except $\sigma_{\beta m2}$
are the same as those in Eq.\,(\ref{sys1})
in the analysis of SK data.
$\sigma_{\beta m2}=0.16$, which
is the uncertainty in the relative normalization between
the $\nu$ - $\bar{\nu}$ flux, was chosen
because this value was used
in the energy spectrum analysis
on the significance of the wrong mass hierarchy
to give the result close to that in Ref.\,\cite{Abe:2011ts}.
(See the discussions below.)

In the spectrum analysis, on the other hand,
$\chi^2_{\rm sub-GeV}$ and
$\chi_{\rm multi-GeV}^2$ are replaced by
\hglue -2cm
\begin{eqnarray}
\hspace*{-20mm}
&{\ }&
\displaystyle\chi_{\rm sub-GeV}^2
\nonumber\\
&=&
\min_{\alpha_s,\beta's,\gamma's}\left[
\frac{\beta_{s1}^2}{\sigma_{\beta s1}^2}
+\frac{\beta_{s2}^2}{\sigma_{\beta s2}^2}
+\frac{\gamma_{L1}^2}{\sigma_{\gamma L1}^2}
+\frac{\gamma_{L2}^2}{\sigma_{\gamma L2}^2}
+\frac{\gamma_{H1}^2}{\sigma_{\gamma H1}^2}
+\frac{\gamma_{H2}^2}{\sigma_{\gamma H2}^2}
\right.
\nonumber\\
&{\ }&\quad
+\sum_{A=L,H}\sum_{j=1}^{10}\left\{
\frac{1}{n_{Aj}^s(e)}
\left[ \alpha_s \left(1-{\beta_{s1} \over 2}+{\beta_{s2}
\over 2} +  {\gamma_{A1}^j \over 2}  \right)N_{Aj}^s(\nu_e\to\nu_e)
\right.\right.
\nonumber\\
&{\ }&\quad
+ \alpha_s \left(1+{\beta_{s1} \over 2}+{\beta_{s2}
\over 2}+{\gamma_{A1}^j \over 2}\right)N_{Aj}^s(\nu_\mu\to\nu_e)
\nonumber\\
&{\ }&\quad+ \alpha_s \left(1-{\beta_{s1} \over 2}-{\beta_{s2}
\over 2}+{\gamma_{A1}^j \over 2}\right)N_{Aj}^s(\bar{\nu}_e\to\bar{\nu}_e)
\nonumber\\
&{\ }&\quad\left.
+ \alpha_s \left(1+{\beta_{s1} \over 2}-{\beta_{s2}
\over 2}+{\gamma_{A1}^j \over 2}\right)N_{Aj}^s(\bar{\nu}_\mu\to\bar{\nu}_e)
-n_{Aj}^s(e)\right]^2
\nonumber\\
&{\ }&\quad
+\frac{1}{n_{Aj}^s(\mu)} \left[ 
\alpha_s \left(1-{\beta_{s1} \over 2}+{\beta_{s2} 
\over 2}+{\gamma_{A2}^j \over 2}\right)N_{Aj}^s(\nu_e\to\nu_\mu)
\right.\nonumber\\
&{\ }&\quad
+\alpha_s \left(1+{\beta_{s1} \over 2}+{\beta_{s2} 
\over 2}+{\gamma_{A2}^j \over 2}\right)N_{Aj}^s(\nu_\mu\to\nu_\mu)
\nonumber\\
&{\ }&\quad+\alpha_s \left(1-{\beta_{s1} \over 2}-{\beta_{s2} 
\over 2}+{\gamma_{A2}^j \over 2}\right)N_{Aj}^s(\bar{\nu}_e\to\bar{\nu}_\mu)
\nonumber\\
&{\ }&\quad\left.\left.\left.+\alpha_s \left(1+{\beta_{s1} \over 2}-{\beta_{s2} 
\over 2}+{\gamma_{A2}^j \over 2}\right)N_{Aj}^s(\bar{\nu}_\mu\to\bar{\nu}_\mu)
-n_{Aj}^s(\mu)\right]^2
\right\}\right],
\label{eqn:chi-sub2}
\end{eqnarray}
\begin{eqnarray}
&{\ }&\displaystyle\chi_{\rm multi-GeV}^2\nonumber\\
&=&
\min_{\alpha_m,\beta's,\gamma's}\left[
\frac{\beta_{m1}^2}{\sigma_{\beta m1}^2}
+\frac{\beta_{m2}^2}{\sigma_{\beta m2}^2}
+\frac{\gamma_{1}^2}{\sigma_{\gamma 1}^2}
+\frac{\gamma_{2}^2}{\sigma_{\gamma 2}^2}\right.
\nonumber\\
&{\ }&+\sum_{A=L,H}\sum_{j=1}^{10}\left\{
\frac{1}{n_{Aj}^m(e)}
\left[ \alpha_s \left(1-{\beta_{m1} \over 2}+{\beta_{m2}
\over 2}+{\gamma_{1}^j \over 2} \right)N_{Aj}^m(\nu_e\to\nu_e)
\right.\right.
\nonumber\\
&{\ }&+ \alpha_s \left(1+{\beta_{m1} \over 2}+{\beta_{m2}
\over 2}+{\gamma_{1}^j \over 2}\right)N_{Aj}^m(\nu_\mu\to\nu_e)
\nonumber\\
&{\ }&+ \alpha_s \left(1-{\beta_{m1} \over 2}-{\beta_{m2}
\over 2}+{\gamma_{1}^j \over 2}\right)N_{Aj}^m(\bar{\nu}_e\to\bar{\nu}_e)
\nonumber\\
&{\ }&\left.+ \alpha_s \left(1+{\beta_{m1} \over 2}-{\beta_{m2}
\over 2}+{\gamma_{1}^j \over 2}\right)N_{Aj}^m(\bar{\nu}_\mu\to\bar{\nu}_e)
-n_{Aj}^m(e)\right]^2
\nonumber\\
&{\ }&+\frac{1}{n_{Aj}^m(\mu)}
\left[ 
\alpha_s \left(1-{\beta_{m1} \over 2}+{\beta_{m2} 
\over 2}+{\gamma_{2}^j \over 2}\right)N_{Aj}^m(\nu_e\to\nu_\mu)\right.
\nonumber\\
&{\ }&+\alpha_s \left(1+{\beta_{m1} \over 2}+{\beta_{m2} 
\over 2}+{\gamma_{2}^j \over 2}\right)N_{Aj}^m(\nu_\mu\to\nu_\mu)
\nonumber\\
&{\ }&+\alpha_s \left(1-{\beta_{m1} \over 2}-{\beta_{m2} 
\over 2}+{\gamma_{2}^j \over 2}\right)N_{Aj}^m(\bar{\nu}_e\to\bar{\nu}_\mu)
\nonumber\\
&{\ }&\left.\left.\left.+\alpha_s \left(1+{\beta_{m1} \over 2}-{\beta_{m2} 
\over 2}+{\gamma_{2}^j \over 2}\right)N_{Aj}^m(\bar{\nu}_\mu\to\bar{\nu}_\mu)
-n_{Aj}^m(\mu)\right]^2
\right\}\right].
\label{eqn:chi-multi2}
\end{eqnarray}
In Eq.\,(\ref{eqn:chi-multi2})
we have introduced the relative normalization,
which in general depends on the flavor
and the energy of the events,
between the upward and downward going bins:
\begin{eqnarray}
\gamma_{A1,2}^j&=&
\left\{ \begin{array}{ll}
\gamma_{A1,2} & (j \le j_{\text{th}}; A=L, H) \nonumber\\
-\gamma_{A1,2} & (j > j_{\text{th}}; A=L, H) \nonumber\\
\end{array} \right.\nonumber\\
\gamma_{1,2}^j&=&
\left\{ \begin{array}{ll}
\gamma_{1,2} & (j \le j_{\text{th}}) \nonumber\\
-\gamma_{1,2} & (j > j_{\text{th}}), \nonumber\\
\end{array} \right.
\end{eqnarray}
and $j_{\text{th}}=3$ is the index which
separates the upward and downward bins.
The indices $L$ and $H$ stand for the
lower ($E<E_{\text{th}}$) and
higher ($E>E_{\text{th}}$) energy bins, 
and the threshold energy $E_{\text{th}}$ is 
chosen so that the numbers of events
for the lower and higher energy bins
are approximately equal, and 
in the case of the sub-GeV events,
$E_{\text{th}}$ = 0.5GeV, and in the
case of the multi-GeV events,
the threshold energy is $E_{\text{th}}$ = 3.2GeV,
respectively, for all the zenith angle bins.
We have put the systematic errors as follows:
\hspace*{-10mm}
\begin{eqnarray}
\hspace*{-10mm}
&{\ }&\sigma_{\beta s1}=\sigma_{\beta m1} = 0.03,
\sigma_{\beta s2} = 0.05,
\sigma_{\beta m2} = 0.16,
\sigma_{\alpha} = 0.2,
\label{sys2}\\
&{\ }&\sigma_{\gamma L1}=0.005, \sigma_{\gamma L2}=0.008,
\sigma_{\gamma H1}=0.021, \sigma_{\gamma H2}=0.018,
\sigma_{\gamma 1}=0.015, \sigma_{\gamma 2}=0.025.  \nonumber\\
\label{sys3}
\end{eqnarray}
All the systematic errors in (\ref{sys2})
except $\sigma_{\beta m2}$ and
$\sigma_{\gamma 2}$
are the same as those in (\ref{sys1})
in sect.\,\ref{atmospheric-sk} and
those used in Ref.\,\cite{Ashie:2005ik}.
$\sigma_{\beta m2}=0.16$
is the uncertainty in the relative normalization between
the multi-GeV $\nu$ - $\bar{\nu}$ flux and it was 0.05 in
(\ref{sys1}).
$\sigma_{\gamma 2}=0.025$
is the uncertainty in the relative normalization between
the upward and downward going multi-GeV $\mu$-like events
and it was 0.008 in the analysis of SK data\,\cite{Ashie:2005ik}.
The choice of these systematic errors (\ref{sys2}) and
(\ref{sys3}) and the index $j_{\text{th}}=3$
has been made
so that the result of our analysis on the mass hierarchy
is close to that in Ref.\,\cite{Abe:2011ts},
and we have checked that
it dose not affect the sensitivity to NSI significantly.

\begin{figure}
\includegraphics[scale=0.4]{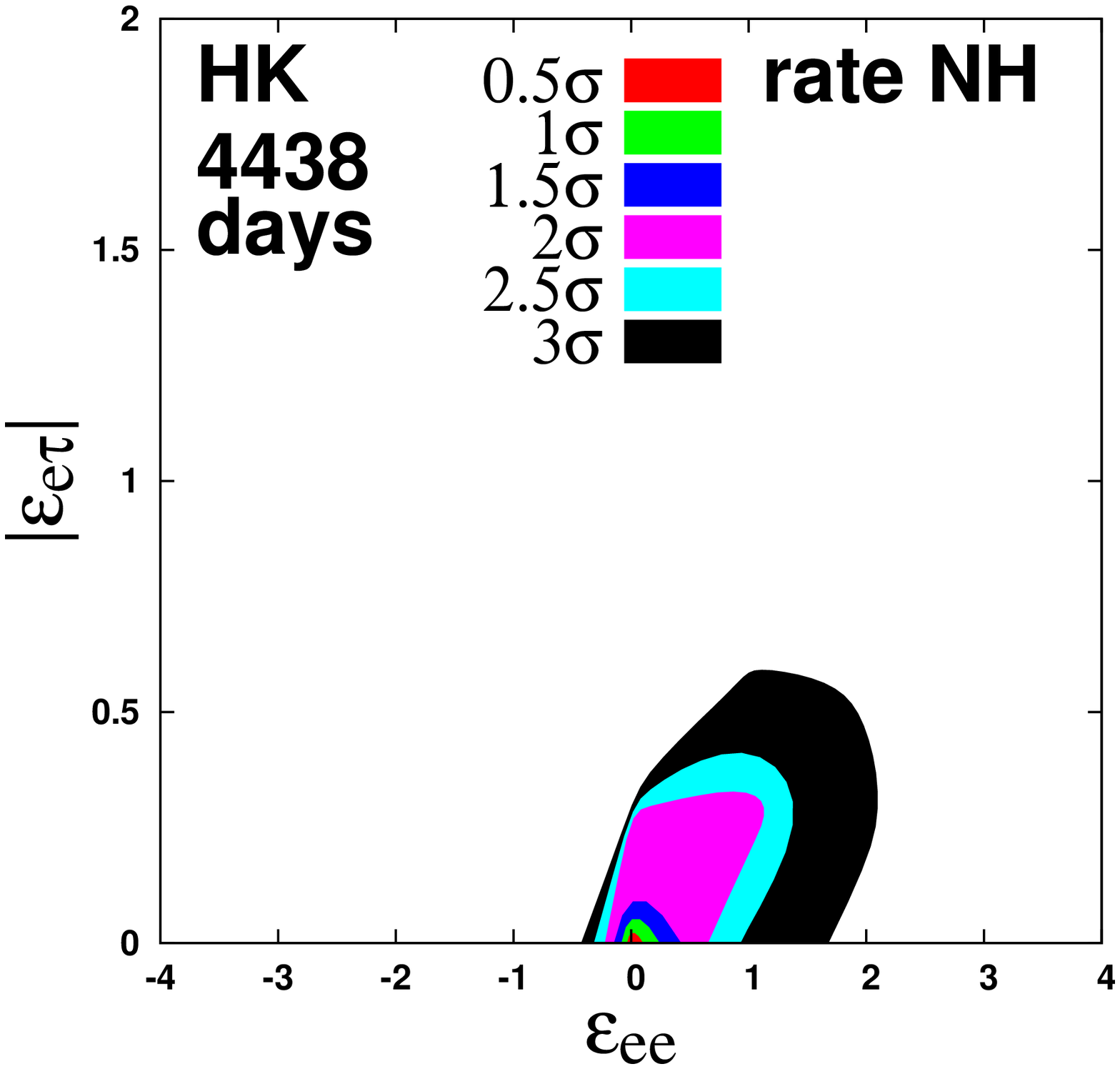}
\includegraphics[scale=0.4]{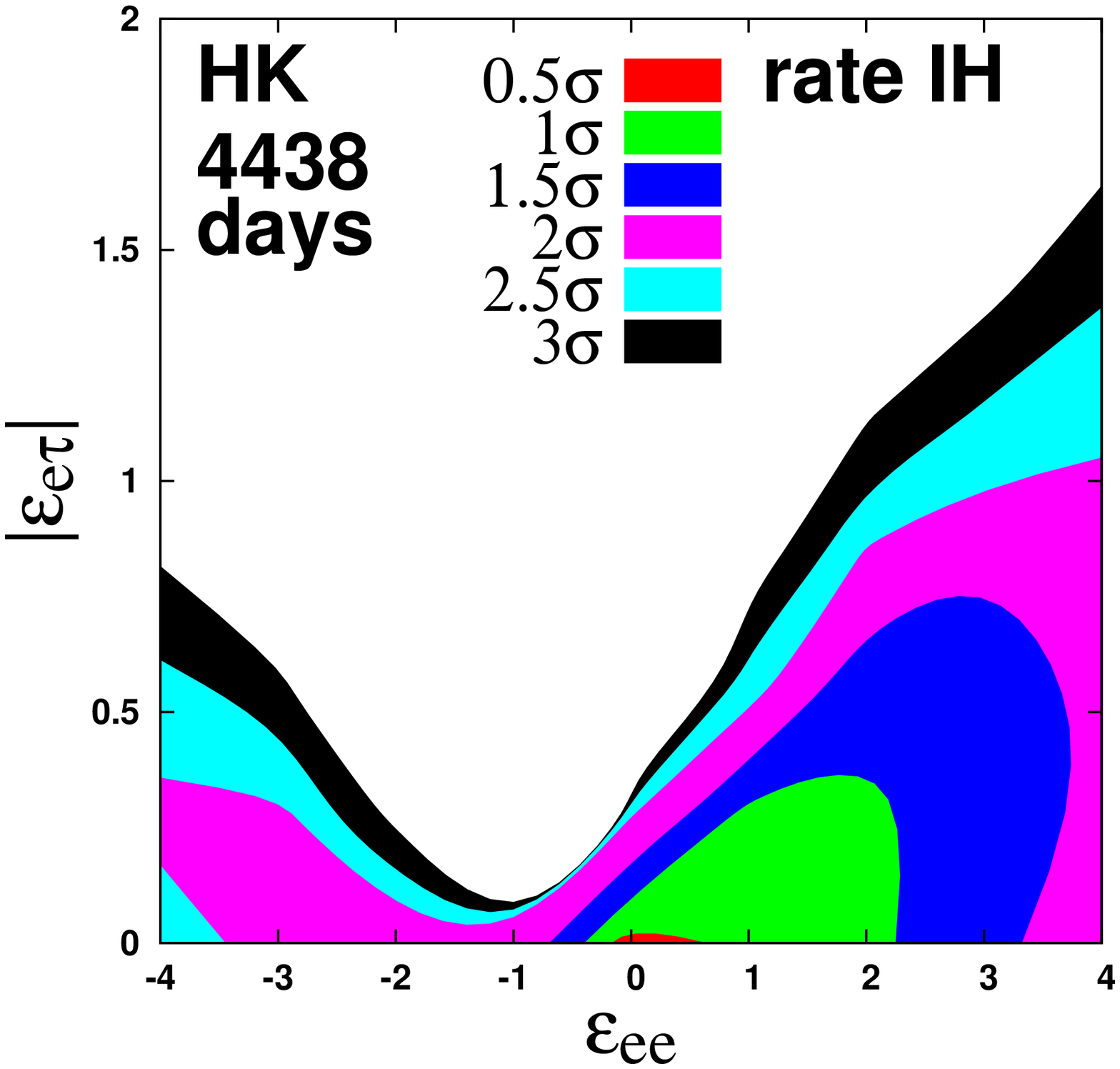}
\includegraphics[scale=0.4]{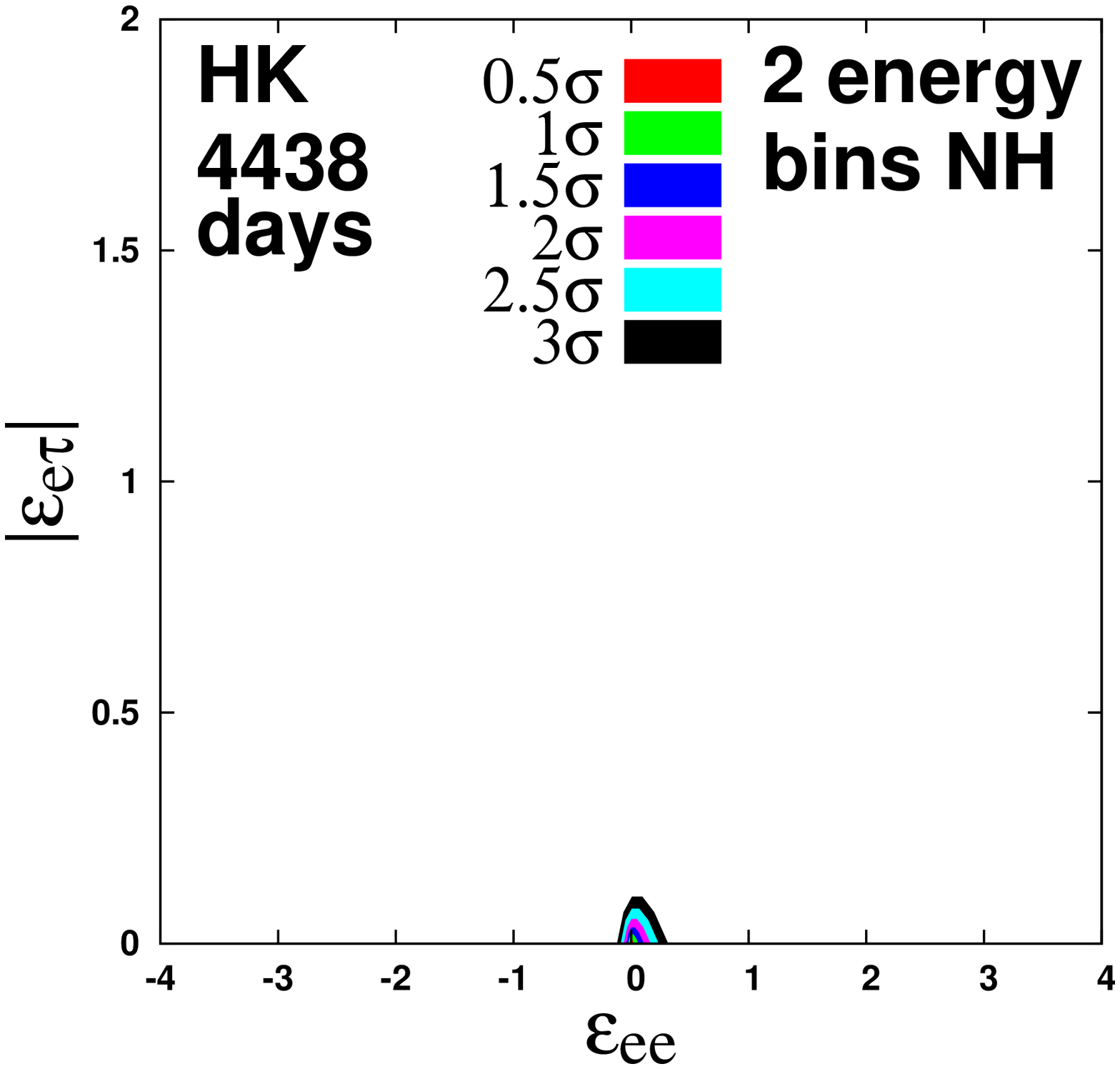}
\includegraphics[scale=0.4]{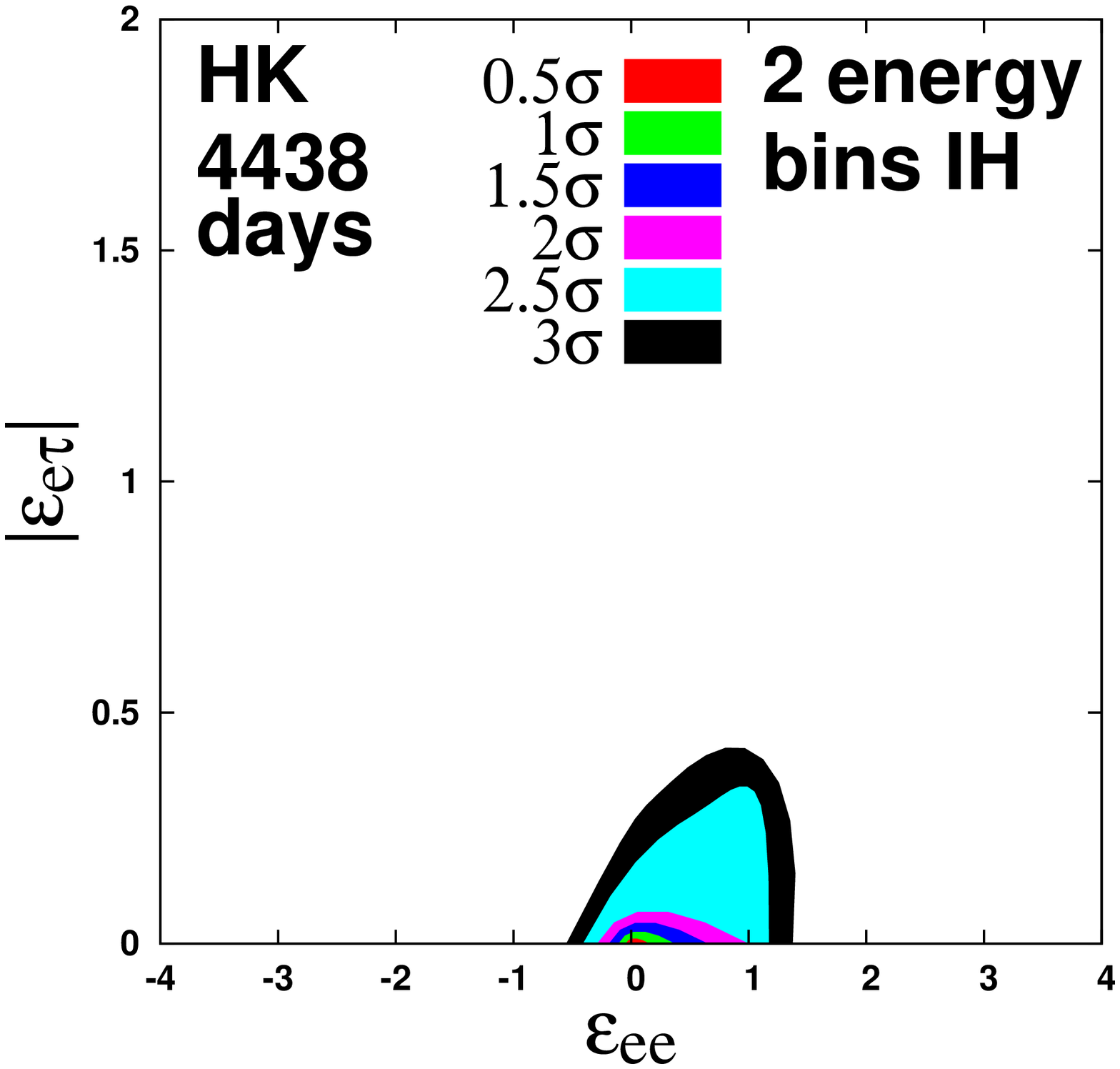}
\vspace*{-5mm}
\caption{
Upper panel: The allowed region
in the ($\epsilon_{ee}$, $|\epsilon_{e\tau}|$) plane
from the HK atmospheric neutrino data for a
normal mass hierarchy (left panel) and for an inverted mass
hierarchy (right panel) from the energy-rate analysis.
Lower panel: The same allowed region as the upper panel
from the two energy-bin analysis.
}
\label{fig:fig2}
\end{figure}

The results from the energy rate
(spectrum)
analysis are given by the upper
(lower) panel in Fig.\,\ref{fig:fig2}.
From the energy rate analysis we have
$|\epsilon_{e\tau}/(1+\epsilon_{ee})|\lesssim 0.3$
at 2.5$\sigma$CL.
On the other hand, from the energy spectrum analysis
we get $-0.1\lesssim \epsilon_{ee}\lesssim 0.2$
and $|\epsilon_{e\tau}|< 0.08$
at 2.5$\sigma$ (98.8\%) CL for the normal hierarchy and
to $-0.4 \lesssim \epsilon_{ee}\lesssim 1.2$
and $|\epsilon_{e\tau}|< 0.34$
at 2.5$\sigma$ (98.8\%) CL for the inverted hierarchy.

From Fig.\,\ref{fig:fig2} we note two things.
Firstly, the allowed regions
from the energy spectrum analysis
(the lower panel)
are much smaller than those
from the energy rate analysis
(the upper panel)
for both mass hierarchies.
Secondly, the allowed regions (the right panel)
for the inverted hierarchy
are wider than those
(the left panel)
for the normal hierarchy
for both rate and spectrum analyses.

\begin{figure}[H]
\includegraphics[scale=0.25]{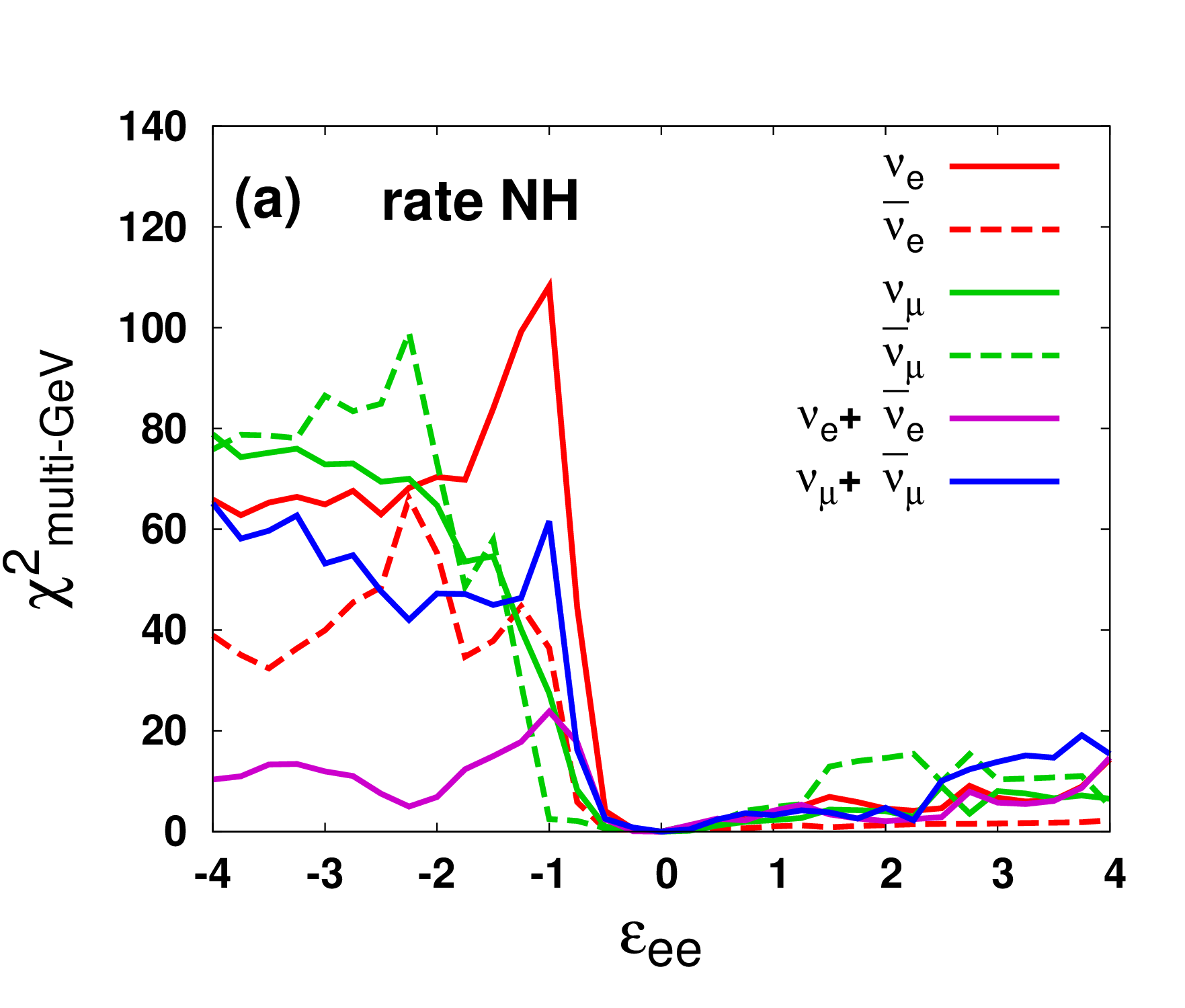}
\includegraphics[scale=0.25]{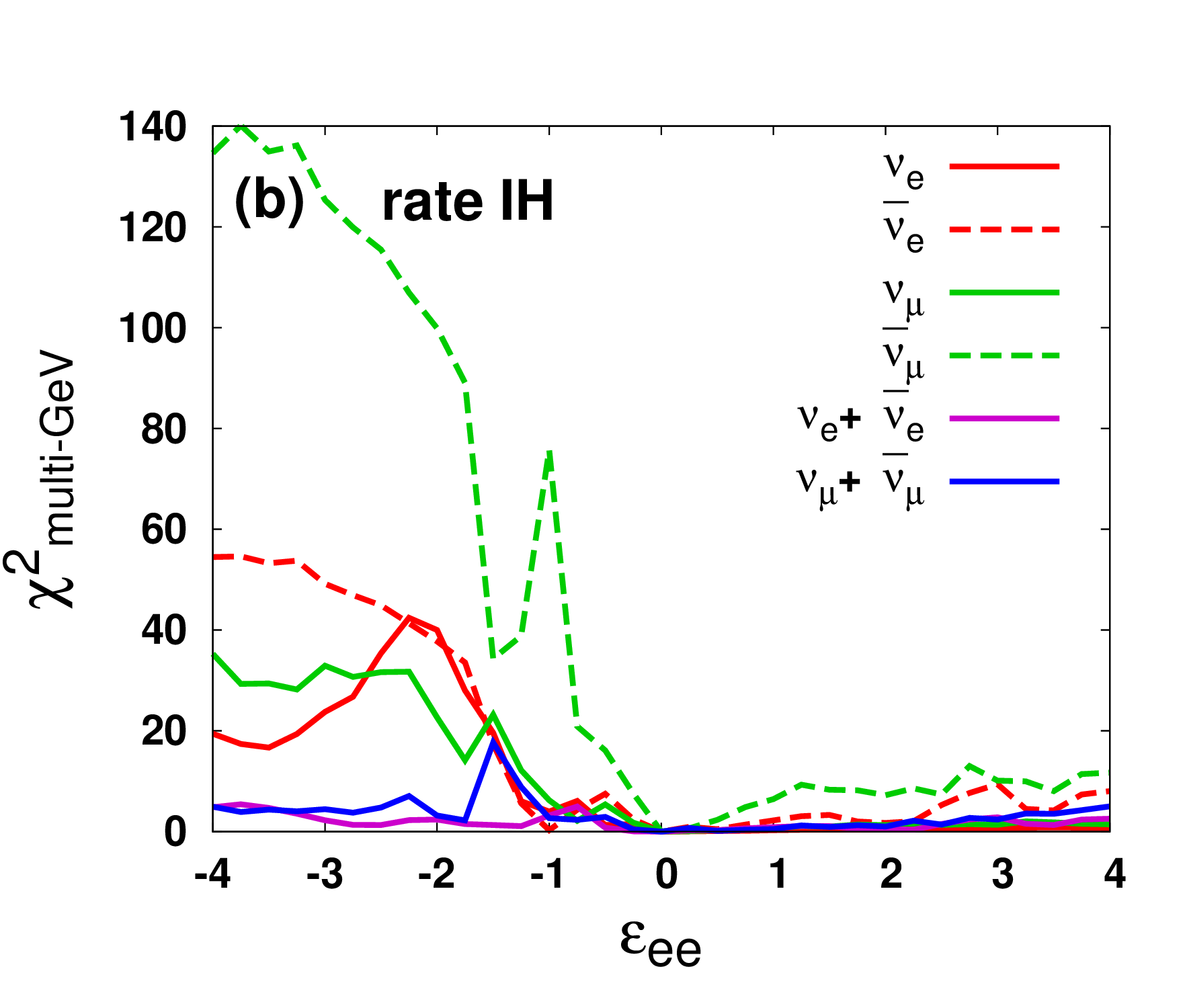}
\includegraphics[scale=0.25]{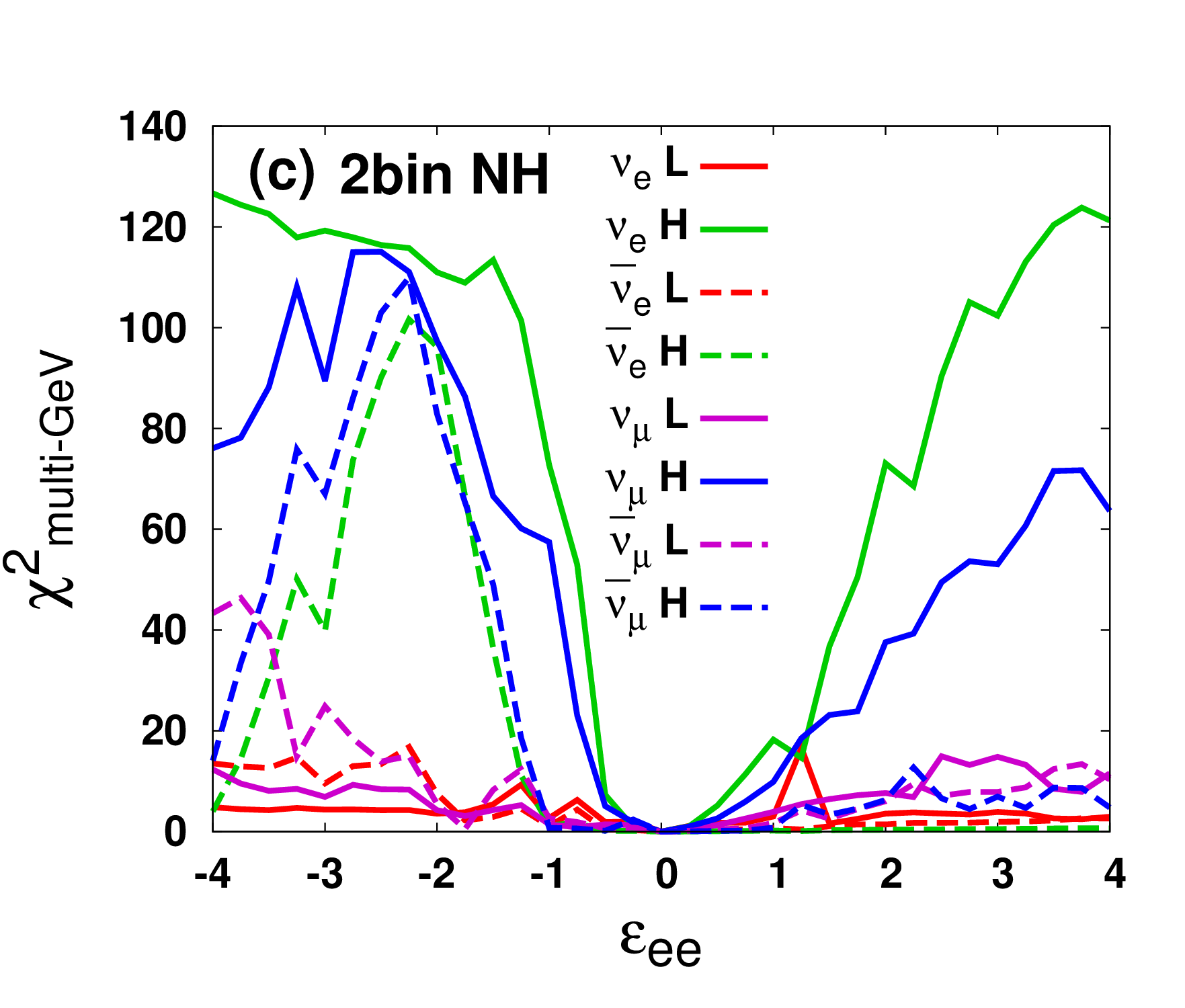}
\includegraphics[scale=0.25]{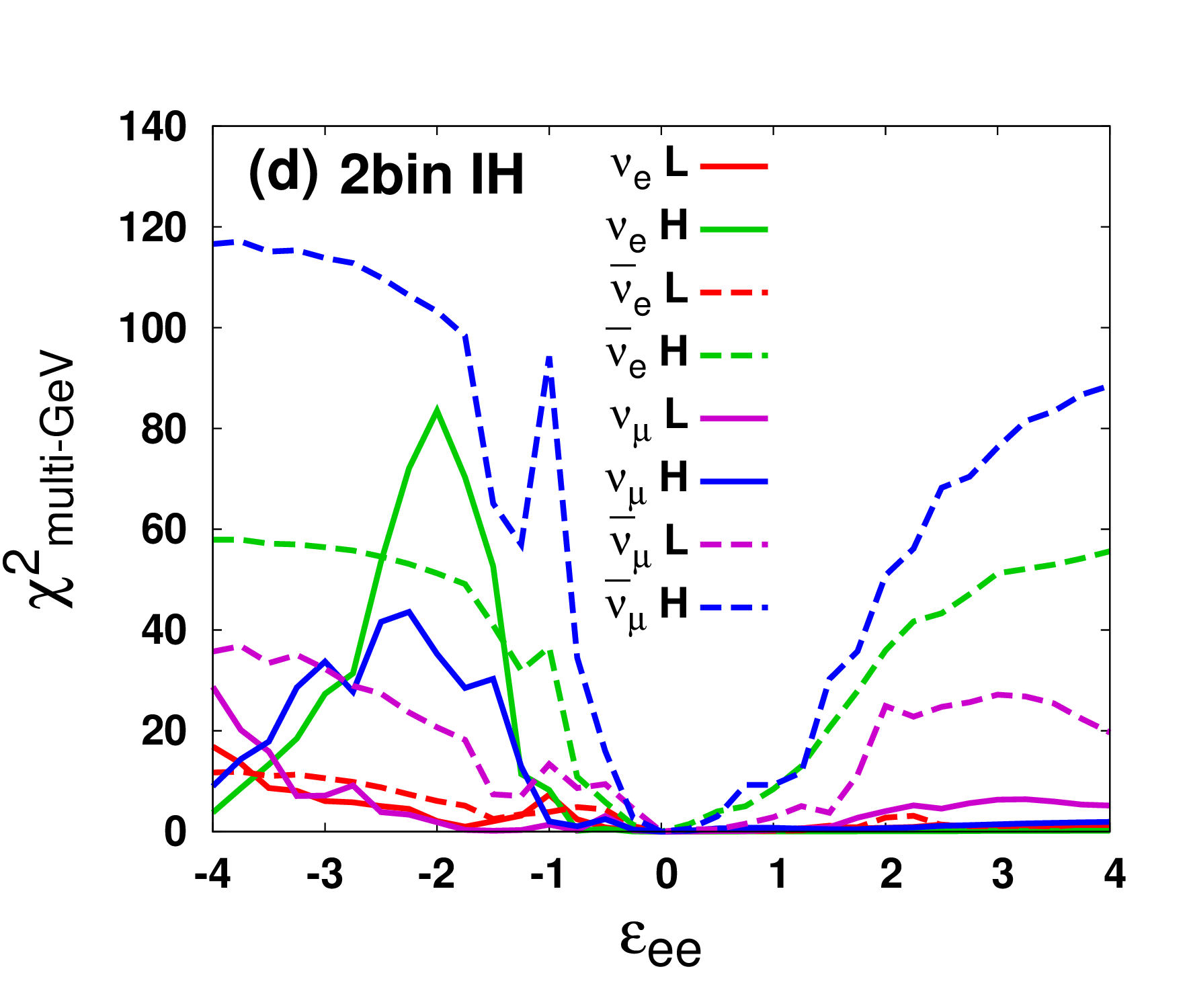}
\includegraphics[scale=0.25]{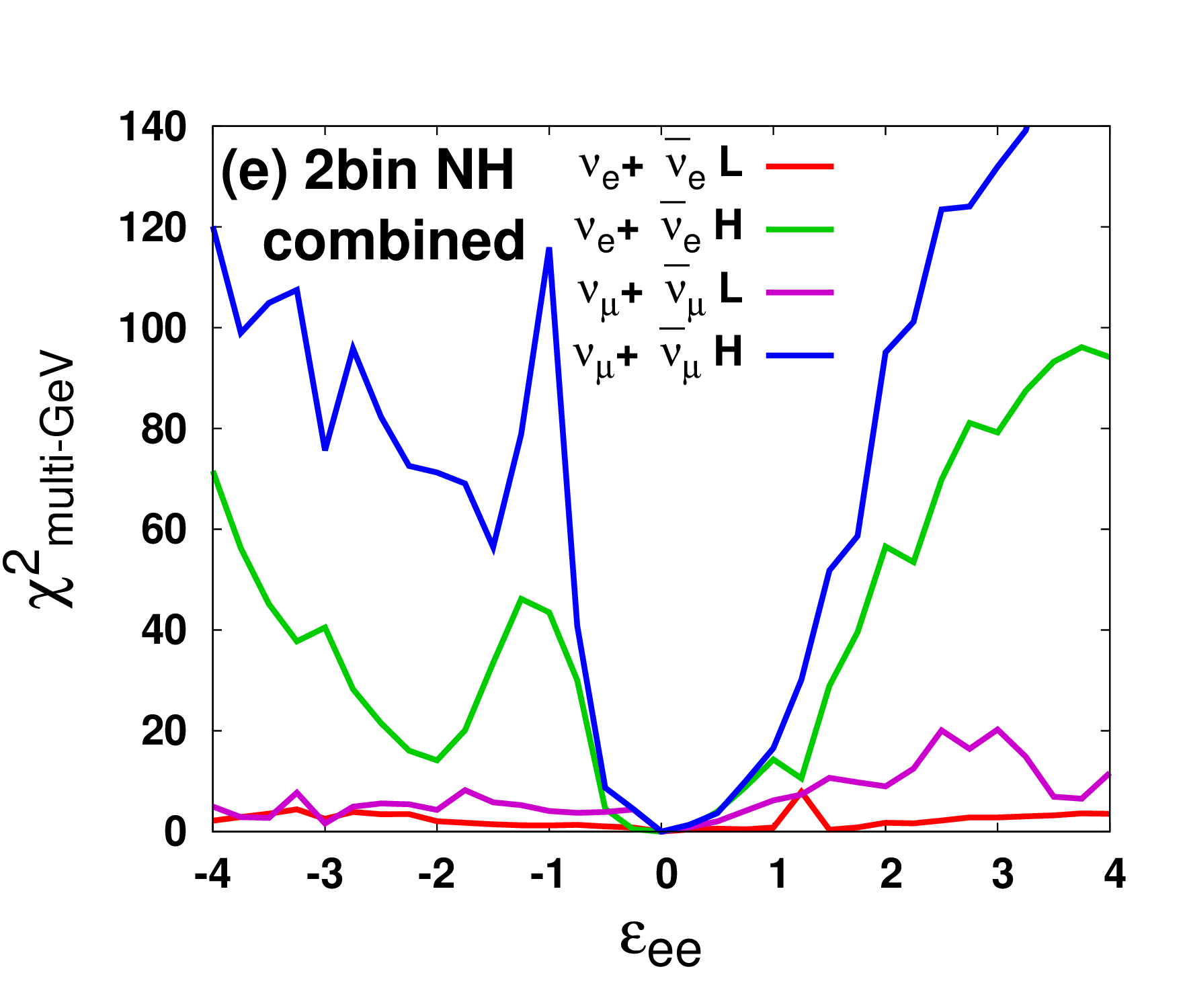}
\includegraphics[scale=0.25]{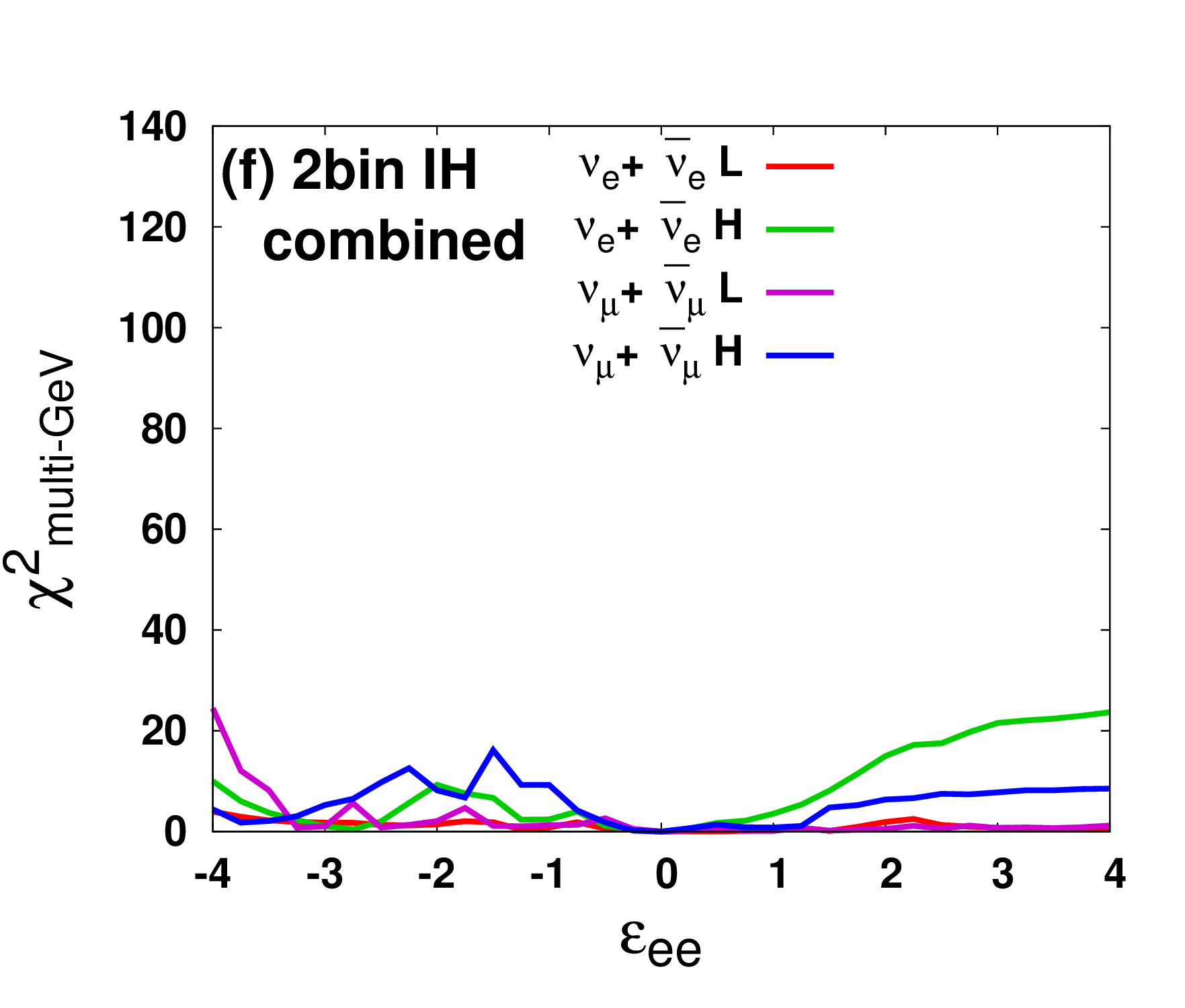}
\caption{
The behaviors of $\chi_{\rm multi-GeV}^2$
for $\epsilon_{e\tau}=0$ as a function of $\epsilon_{ee}$.
(a), (b): Energy rate analysis for NH (a) and IH (b).
(c), (d): Energy spectrum analysis for NH (d) and IH (e)
for the separate neutrino or antineutrino events.
(e), (f): Energy spectrum analysis for NH (e) and IH (f)
using only the combined numbers of events
of $\nu_e+\bar{\nu}_e$ and $\nu_\mu+\bar{\nu}_\mu$.
In (a), (b), (c) and (d), the plots for
the separate neutrino or antineutrino events
are created based on the assumption that HK could distinguish
neutrinos and antineutrinos.
}
\label{fig:fig3}
\end{figure}

To understand these phenomena,
we have plotted in Fig.\,\ref{fig:fig3}
$\chi_{\rm multi-GeV}^2$
for the multi-GeV events, which are
expected to be sensitive to the
matter effect and therefore to $\epsilon_{ee}$,
as a function of $\epsilon_{ee}$
in the case of $\epsilon_{e\tau}=0$.
In plotting the figures in Fig.\,\ref{fig:fig3},
we have taken into account only the statistical
errors for simplicity, and
we assume that the HK detector could distinguish
neutrinos and antineutrinos for both $e$-like
and $\mu$-like events in all the energy
ranges of the multi-GeV events, and that
the detection efficiency
is the same for both neutrinos and antineutrinos.
Since the SK collaboration
distinguish neutrinos and antineutrinos only
for the multi-GeV e-like events\,\cite{Abe:2014gda},
our assumption here may not be realistic, and the
separate plots for neutrinos or for antineutrinos
except for the e-like events
should be regarded as information for
theoretical consideration.
The two figures ((a) and (b)) in the top row
are the results of the energy rate analysis.
The two figures ((c) and (d)) in the middle row
are the results of the energy spectrum
analysis with two energy bins for the
separate neutrino or antineutrino events.
The two figures ((e) and (f)) in the bottom row
are the results of the energy spectrum
analysis with two energy bins of
neutrinos and antineutrinos combined.
Comparing the figures ((a) and (b)) in the top row
and those ((e) and (f)) in the bottom row, we see that,
even if some of the data set in the
spectrum analysis have a sensitivity to
the effect of $\epsilon_{ee}$,
the data in the rate analysis does not necessarily
have a sensitivity to $\epsilon_{ee}$
particularly for $\epsilon_{ee}>0$,
for both mass hierarchies.
While it is not clear to us why the
sensitivity is lost only for
$\epsilon_{ee}>0$, we have found that,
if we try to fit the same data with
the numbers of events with the wrong
mass hierarchy, then the
plot becomes left-right reversed,
i.e., the sensitivity is lost only for
$\epsilon_{ee}<0$.
On the other hand,
by  comparing the figures ((c) and (d)) in the middle row
and those ((e) and (f)) in the bottom row, we see that,
in the case of the inverted mass hierarchy,
even though the separate $\bar{\nu}_\mu$ data
has a sensitivity to $\epsilon_{ee}$,
the combined data $\nu_\mu+\bar{\nu}_\mu$
loses a sensitivity.
We could not explain these phenomena
using the analytic expression for the
oscillation probability, but we interpret
this loss of sensitivity as a
destructive phenomenon
between neutrinos and antineutrinos
in the rate analysis, and between the
lower and higher energy bins
in the spectrum analysis for
the inverted mass hierarchy.

It is expected that the HK experiment will
be able to use information on the energy spectrum,
so we believe that the allowed region in the lower panel
in Fig.\,\ref{fig:fig2} with the energy
spectrum analysis reflects the true HK sensitivity
more than that in the upper panel does.

\subsection{The case  in the presence of NSI
\label{hk-nsi}}

\begin{figure}[H]
\includegraphics[width=.5\textwidth]{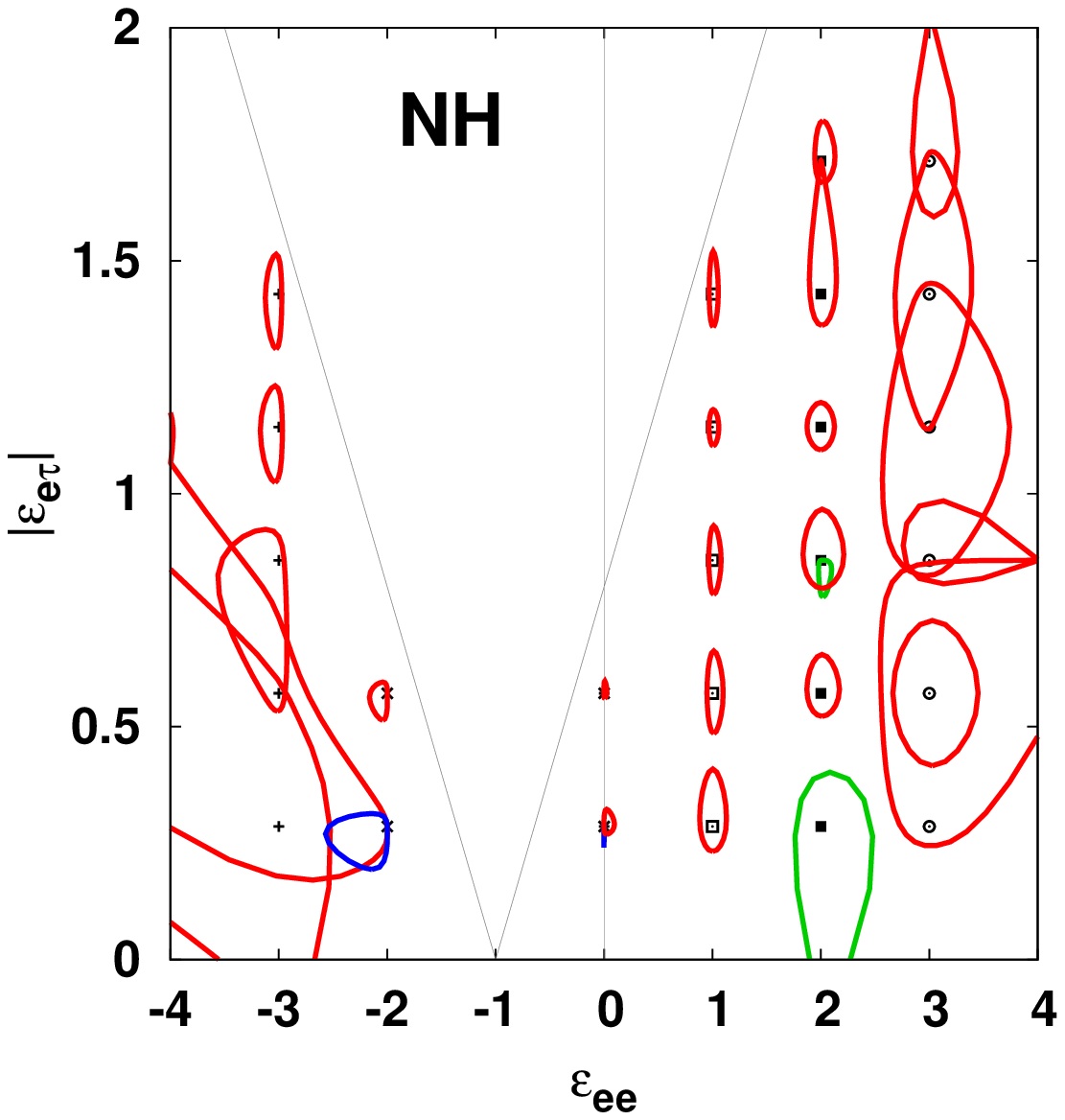}
\hspace*{-3mm}
\includegraphics[width=.5\textwidth]{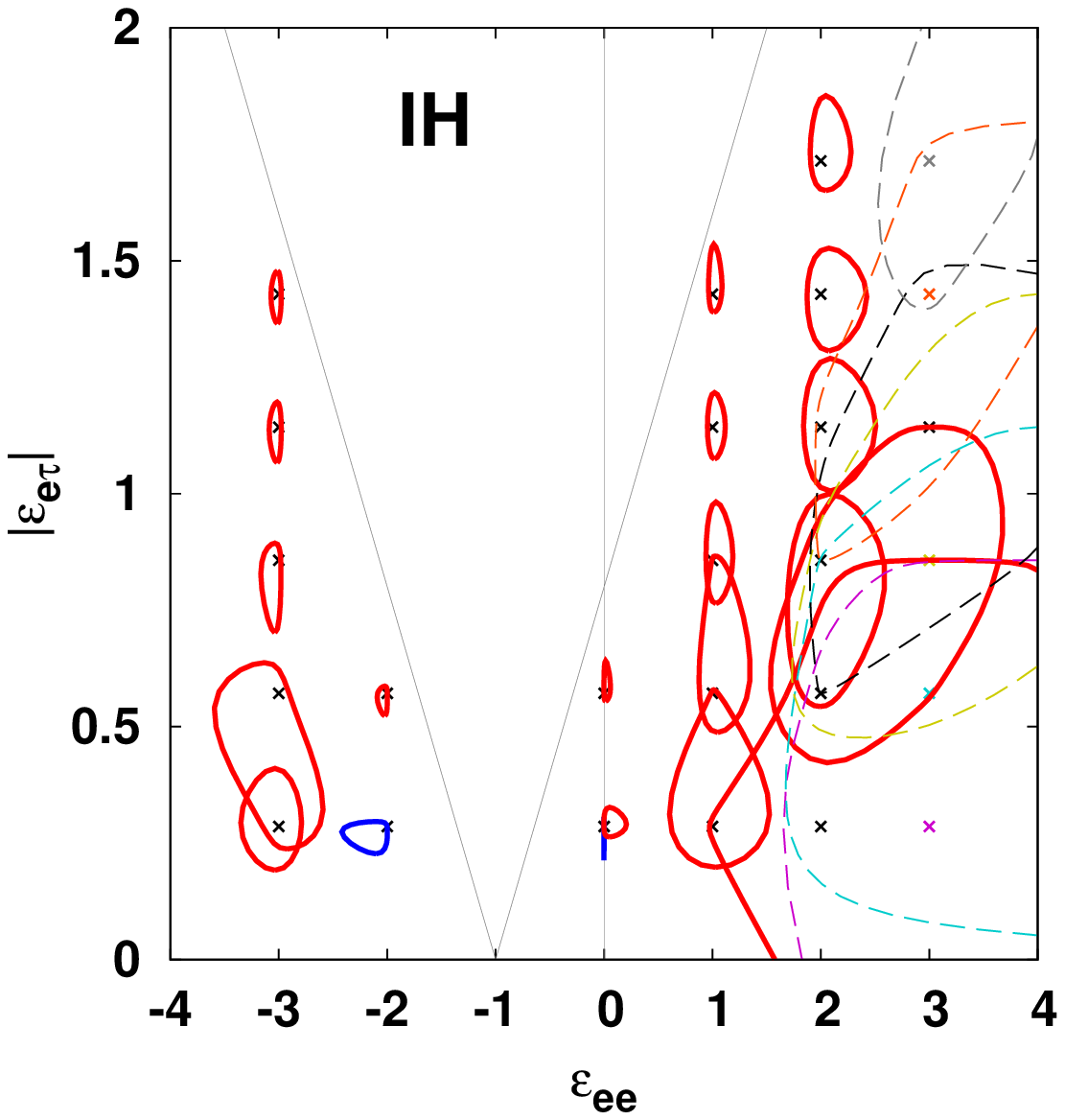}
\caption{The allowed region at 2.5$\sigma$CL around the point
$(\epsilon_{ee}, |\epsilon_{e\tau}|) \ne (0,0)$,
where $\bar{\delta}=\mbox{\rm arg}(\bar{\epsilon}_{e\tau})=0$ is assumed.
Most of the allowed regions are connected, but
those around a few points have an isolated island, and they
are depicted with different colors.
The allowed regions at $\epsilon_{ee}=3$ for
the inverted mass hierarchy are much wider
compared with other cases, so their boundary
as well as their center are shown
with dashed lines and with different colors.}
\label{fig:fig4}
\end{figure}

Next let us discuss the case where
``the experimental data'' is the one
obtained with $(\epsilon_{ee}, \epsilon_{e\tau}) \ne (0,0)$.
The analysis is the same as the one in subsect.\,\ref{hk-std},
except that the ``the experimental data'' is produced
assuming the presence of NSI, and here
we perform only an energy spectrum analysis with
two energy bins.
The results are shown in Fig.\,\ref{fig:fig4}, where
the allowed regions at 2.5$\sigma$CL
($\Delta\chi^2=8.8$ for 2 degrees of freedom)
around the true points are depicted.
The straight lines $|\epsilon_{e\tau}|=0.8\times|1+\epsilon_{ee}|$
in Fig.\,\ref{fig:fig4} stand for the approximate bound
from the SK atmospheric neutrinos in  Fig.\,\ref{fig:fig1},
and we have examined only the points below
these straight lines.
As seen from Fig.\,\ref{fig:fig4}, 
the errors in $\epsilon_{ee}$ and
$|\epsilon_{e\tau}|$ are small
for $|\epsilon_{ee}| \lesssim 2$
in the case of the normal hierarchy
and for $-3 \lesssim \epsilon_{ee} \lesssim 1$
in the case of the inverted hierarchy.
The errors are large otherwise,
and the reason that the errors are large
is because a sensitivity is lost due to a
destructive phenomenon
between neutrinos and antineutrinos
as was discussed in subsect.\,\ref{hk-std}.

We note in passing that there are a couple of
points in Fig.\,\ref{fig:fig4}, where
the allowed region has an additional isolated island.
This is regarded as so-called parameter
degeneracy\,\cite{BurguetCastell:2001ez,Minakata:2001qm,Fogli:1996pv,Barger:2001yr}
in the presence of the NSI.
Since little is known about parameter degeneracy
in the presence of the new physics
and since the study of the subject is beyond the scope of this paper,
we do not discuss parameter degeneracy here.

\section{Conclusions
\label{conclusions}}

In this paper we have studied the constraint
of the SK atmospheric neutrino data on
the non-standard flavor-dependent
interaction in neutrino propagation with the
ansatz (\ref{ansatz}).
From the SK atmospheric neutrino data for 4438 days,
we have obtained the bound
$|\epsilon_{e\tau}|/|1+\epsilon_{ee}|\lesssim 0.8$
at 2.5$\sigma$CL,
while we have little constraint on $\epsilon_{ee}$.

We have also discussed
the sensitivity of the future HK atmospheric neutrino
experiment to NSI by analyses with the energy rate
and with the energy spectrum.
If nature is described by the standard oscillation
scenario, then the HK atmospheric neutrino data
will give us the bound
$|\epsilon_{e\tau}|/|1+\epsilon_{ee}|\lesssim 0.3$
at 2.5$\sigma$CL from the energy rate analysis, and
from the energy spectrum analysis
it will restrict $\epsilon_{ee}$ to
$-0.1\lesssim \epsilon_{ee}\lesssim 0.2$
and $|\epsilon_{e\tau}|< 0.08$
at 2.5$\sigma$ (98.8\%) CL for the normal hierarchy and
to $-0.4 \lesssim \epsilon_{ee}\lesssim 1.2$
and $|\epsilon_{e\tau}|< 0.34$
at 2.5$\sigma$ (98.8\%) CL for the inverted hierarchy.
On the other hand, if nature is described by
NSI with the ansatz (\ref{ansatz}), then
HK will measure the NSI parameters
$\epsilon_{ee}$ and $|\epsilon_{e\tau}|$
relatively well
for $|\epsilon_{ee}| \lesssim 2$
in the case of the normal hierarchy
and for $-3 \lesssim \epsilon_{ee} \lesssim 1$
in the case of the inverted hierarchy.

We have shown that it is important to use
information on the energy spectrum
to obtain strong constraint,
because a sensitivity to NSI
would be lost due to
a destructive phenomena between
the low and high energy events.
If there is a way
to distinguish between neutrinos and antineutrinos,
as is done by the SK collaboration\,\cite{Abe:2014gda}
for the e-like multi-GeV events,
also for the multi-GeV $\mu$-like events,
then the sensitivity to NSI
would be greatly improved,
because in this case we can avoid
a destructive phenomena between
neutrinos and antineutrinos.

While HK is expected to play an important role
in measurement of $\delta$ in the standard
three-flavor scenario using the JPARC beam,
HK has also a potential for new physics
with atmospheric neutrinos.
Search for NSI may lead to physics beyond the Standard Model,
and the effects of NSI at HK deserves further studies.

\section*{Acknowledgments}
This research was partly supported by a Grant-in-Aid for Scientific
Research of the Ministry of Education, Science and Culture, under
Grants No. 24540281 and No. 25105009.

\end{document}